\documentclass[useAMS,usenatbib]{mn2e}
\usepackage{graphicx}

\newcommand{\ltarget}{SDSS\,J085746.18+034255.3}
\newcommand{\ctarget}{CSS\,03170}
\newcommand{\target}{SDSS\,J0857+0342}

\title[The Eclipsing binary SDSS\,J085746.18+034255.3]{The shortest period detached white dwarf + main-sequence binary}

\author[S. G. Parsons et al.]{S.~G.~Parsons$^{1}$\thanks{steven.parsons@warwick.ac.uk},
T.~R.~Marsh$^{1}$,
B.~T.~G{\"a}nsicke$^{1}$,
V.~S.~Dhillon$^{2}$,
C.~M.~Copperwheat$^{1}$,
\newauthor
S.~P.~Littlefair$^{2}$,
S.~Pyrzas$^{1}$,
A.~J.~Drake$^{3}$,
D.~Koester$^{4}$,
M.~R.~Schreiber$^{5}$
\newauthor
and A.~Rebassa-Mansergas$^{5}$\\
$^{1}$Department of Physics, University of Warwick, Coventry, CV4 7AL\\
$^{2}$Department of Physics and Astronomy, University of Sheffield, Sheffield, S3 7RH \\
$^{3}$California Institute of Technology, 1200 E. California Blvd, CA 91225, USA\\
$^{4}$Institut f{\"u}r Theoretische Physik und Astrophysik, Universit{\"a}t Kiel, Germany\\
$^{5}$Departmento de F{\'i}sica y Astronom{\'i}a, Universidad de Valpara{\'i}so, Avenida Gran Bretana 1111, Valpara{\'i}so, Chile}
\begin{document}
\input{pjw_aas_macros.cls}
\date{Accepted 2011 August 24. Received 2011 August 15; in original form 2011 June 24}

\pagerange{\pageref{firstpage}--\pageref{lastpage}} \pubyear{2011}

\maketitle

\label{firstpage}

\begin{abstract}

We present high-speed ULTRACAM and SOFI photometry and X-shooter spectroscopy of the recently discovered 94 minute orbital period eclipsing white dwarf / main-sequence binary {\ltarget} (\ctarget) and use these observations to measure the system parameters. We detect a shallow secondary eclipse and hence are able to determine an orbital inclination of $i=85.5^{\circ}\pm0.2^{\circ}$. The white dwarf has a mass of $0.51\pm0.05$M$_{\sun}$ and a radius of $0.0247\pm0.0008$R$_{\sun}$. With a temperature of $35,300\pm400$K the white dwarf is highly over-inflated if it has a carbon-oxygen core, however if it has a helium core then its mass and radius are consistent with evolutionary models. Therefore, the white dwarf in {\ltarget} is most likely a helium core white dwarf with a mass close to the upper limit expected from evolution. The main-sequence star is an M8 dwarf with a mass of $0.09\pm0.01$M$_{\sun}$ and a radius of $0.110\pm0.004$R$_{\sun}$ placing it close to the hydrogen burning limit. The system emerged from a common envelope $\sim20$ million years ago and will reach a semi-detached configuration in $\sim400$ million years, becoming a cataclysmic variable with a period of 66 minutes, below the period minimum.

\end{abstract}


\section{Introduction}

Mass-radius relations are of fundamental importance in a wide range of astrophysical circumstances. They are routinely used to infer accurate masses and radii of transiting exoplanets, calibrating stellar evolutionary models and understanding the late evolution of mass transferring binaries such as cataclysmic variables (\citealt{littlefair08}; \citealt{savoury11}). Additionally, the mass-radius relation for white dwarfs has played an important role in estimating the distance to globular clusters \citep{renzini96} and the determination of the age of the Galactic disk \citep{wood92}. Mass-radius relations play a crucial role in determining stellar parameters of single, isolated stars as well as in non-eclipsing binaries, where direct measurements of masses and radii are not possible. Despite their importance, the mass-radius relations for both white dwarfs and low-mass stars remain largely untested.

\begin{table*}
 \centering
  \caption{Journal of observations. Exposure times for X-shooter observations are for UVB arm, VIS arm and NIR arm respectively. The primary eclipse occurs at phase 1, 2 etc.}
  \label{obs}
  \begin{tabular}{@{}lcccccc@{}}
  \hline
  Date at     &Instrument&Filter(s) &Start  &Orbital &Exposure &Conditions             \\
  start of run&          &          &(UT)   &phase   &time (s) &(Transparency, seeing) \\
 \hline
2010/12/03 & ULTRACAM  & $u'g'r'$ & 07:14 & 0.86--1.08 & 4.0  & Excellent, $\sim1$ arcsec \\
2010/12/07 & ULTRACAM  & $u'g'r'$ & 06:53 & 0.95--1.20 & 4.0  & Excellent, $\sim1$ arcsec \\
2010/12/08 & ULTRACAM  & $u'g'r'$ & 06:31 & 0.21--0.63 & 5.0  & Excellent, $\sim1$ arcsec \\
2010/12/09 & ULTRACAM  & $u'g'i'$ & 06:49 & 0.77--1.80 & 5.0  & Excellent, $\sim1$ arcsec \\
2010/12/10 & ULTRACAM  & $u'g'r'$ & 06:31 & 0.93--1.29 & 4.0  & Good, $1.0-1.5$ arcsec    \\
2010/12/16 & ULTRACAM  & $u'g'i'$ & 04:16 & 0.68--2.12 & 4.0  & Good, $1.0-2.0$ arcsec    \\
2011/01/07 & ULTRACAM  & $u'g'i'$ & 06:53 & 0.54--1.11 & 4.0  & Good, $1.0-1.5$ arcsec    \\
2011/01/09 & ULTRACAM  & $u'g'r'$ & 04:46 & 0.70--1.69 & 4.0  & Good, $1.0-1.5$ arcsec    \\
2011/01/10 & ULTRACAM  & $u'g'r'$ & 04:42 & 0.03--0.85 & 4.0  & Good, $1.0-1.5$ arcsec    \\
2011/01/11 & ULTRACAM  & $u'g'r'$ & 03:22 & 0.53--1.95 & 4.0  & Good, $1.0-1.5$ arcsec    \\
2011/02/27 & X-shooter & -        & 03:59 & 0.94--2.14 & 191,213,480 & Good, $1.0-1.5$ arcsec \\
2011/03/03 & X-shooter & -        & 02:15 & 0.27--1.35 & 191,213,480 & Good, $1.0-1.5$ arcsec \\
2011/04/06 & SOFI      & $J$      & 00:58 & 0.72--1.27 & 15.0 & Excellent, $<1$ arcsec    \\
\hline
\end{tabular}
\end{table*}

\citet{provencal98} tested the white dwarf mass-radius relation using \emph{Hipparcos} parallaxes to determine the radii for white dwarfs in visual binaries, common proper-motion (CPM) systems and field white dwarfs. However, the radius measurements for all of these systems still rely to some extent on model atmosphere calculations. For field white dwarfs the mass determinations are also indirect. \citet{barstow05} used {\it Hubble Space Telescope}/STIS spectra to measure the mass of Sirius B to high precision, however, their radius constraint still relied on model atmosphere calculations and is therefore less direct when it comes to testing white dwarf mass-radius relations. To date, only two white dwarfs have had their masses and radii model-independently measured, V471 Tau \citep{obrien01} and NN Ser \citep{parsons10}. Both of these systems are eclipsing post common envelope binaries (PCEBs), demonstrating the importance of eclipsing systems in determining precise masses and radii.

PCEBs are the product of main sequence binary stars where the more massive members of the binaries evolve off the main-sequence and fill their Roche lobes on either the giant or asymptotic giant branches. This initiates a dynamically unstable mass transfer to the less massive components which are unable to accrete the material, hence common envelopes of material are formed containing the core of the giant and the companion star. Frictional forces within this envelope cause the two stars to spiral inwards. The ensuing loss of angular momentum expels the common envelope revealing the tightly bound core and companion. An additional advantage to studying PCEBs is that some contain low-mass helium core white dwarfs, enabling us to measure masses and radii for these white dwarfs which are not found in wide binaries (that did not undergo a common envelope phase).

PCEBs usually contain a low-mass late-type companion to the white dwarf \citep{rebassa11}. Hence an additional benefit of studying PCEBs is that, under favourable circumstances, not only are the white dwarf's mass and radius determined independently of any model, so too are the mass and radius of its companion. There is disagreement between models and observations of low mass stars; the models tend to under predict the radii by as much as 20-30 per cent \citep{lopez07}. However, it is unclear whether this effect is a genuine problem in the models, or a systematic effect caused by testing the mass-radius relation with close binaries, which are tidally locked into rapid rotation and thus can be inflated by stronger magnetic fields inhibiting the efficiency of convection. Recently, \citet{kraus11} analysed 6 new M dwarf eclipsing binary systems and found that the stars in short period binaries ($P \la 1$ day) have radii that are inflated by up to 10 per cent compared to evolutionary models, whilst those in longer period binaries have smaller radii. This supports the inflation via rapid rotation hypothesis. However, their sample contains stars of masses $M \ga 0.4M_{\sun}$ and thus does not test this effect in the very low mass range. Indeed the radius of the rapidly rotating low-mass ($M=0.111M_{\sun}$) M dwarf in the PCEB NN Ser showed very little inflation above evolutionary models, and what inflation there was could be entirely explained as the result of irradiation from its hot white dwarf companion \citep{parsons10}. Furthermore, \citet{berger06} found a similar underestimation of the radius of single stars compared to evolutionary models, implying that rapid rotation may not be the only cause, although they note that their lower metallicity targets do show better agreement with models.

The subject of this paper, {\ltarget} (henceforth \target), was first discovered as a hot ($T_\mathrm{eff}=36181\pm390$K) white dwarf from the Sloan Digital Sky Survey by \citet{eisenstein06}. Regular eclipses were observed by \citet{drake10} as part of the Catalina Real-time Transient Survey. They determined an orbital period of 94 minutes making {\target} the shortest known period, detached, white dwarf / main-sequence binary. 

In this paper we present a combination of ULTRACAM and SOFI high-speed photometry and X-shooter spectroscopy of {\target} and use these to constrain the system parameters. We also discuss the past and future evolution of the system.

\section{Observations and their reduction}

\begin{figure*}
\begin{center}
 \includegraphics[width=0.75\textwidth]{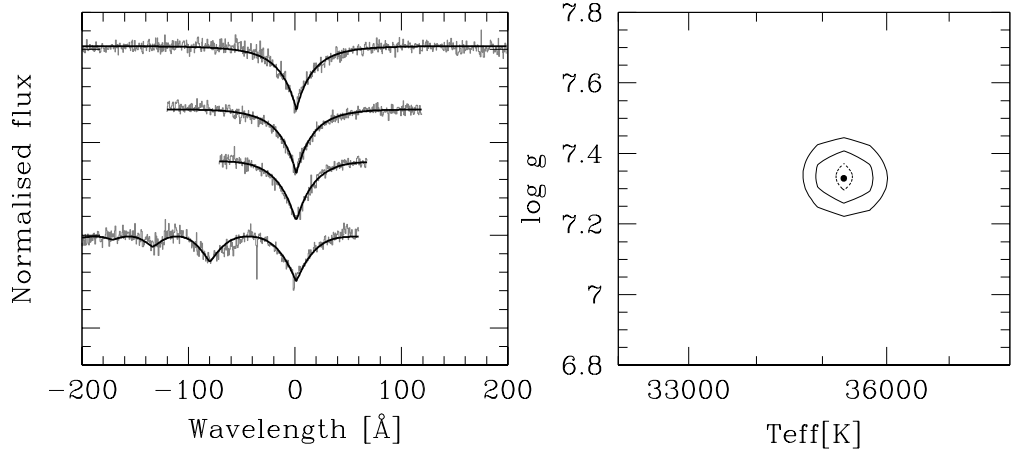}
 \caption{Spectral model fit to the white dwarf component of {\target} from the X-shooter spectroscopy. {\it Left:} best fit (black lines) to the normalised H$\beta$ to H9 line profiles (gray lines, top to bottom). Interstellar calcium absorption is also visible. {\it Right:} 1, 2 and 3$\sigma$ $\chi^2$ contour plots in the $T_\mathrm{eff} - \log{g}$ plane to the line profile fits. The intrinsic contribution from the secondary star in this wavelength range is negligible. We remove the reflection component by only selecting those spectra taken around the primary eclipse, where the reflection is minimal.}
 \label{specfit}
 \end{center}
\end{figure*}

\subsection{NTT/ULTRACAM photometry}

{\target} was observed with ULTRACAM mounted as a visitor instrument on the New Technology Telescope (NTT) at La Silla between December 2010 and January 2011. ULTRACAM is a high-speed, triple-beam CCD camera \citep{dhillon07} which can acquire simultaneous images in three different bands; for our observations we used the SDSS $u'$, $g'$ and either $r'$ or $i'$ filters. A complete log of these observations is given in Table~\ref{obs}. We windowed the CCD in order to achieve exposure times of $\sim$4 seconds which we varied to account for the conditions; the dead time between exposures was $\sim 25$ ms.

All of these data were reduced using the ULTRACAM pipeline software. Debiassing, flatfielding and sky background subtraction were performed in the standard way. The source flux was determined with aperture photometry using a variable aperture, whereby the radius of the aperture is scaled according to the full width at half maximum (FWHM). Variations in observing conditions were accounted for by determining the flux relative to a comparison star in the field of view. The data were flux calibrated by determining atmospheric extinction coefficients in each of the bands in which we observed and we calculated the absolute flux of our target using observations of standard stars (from \citealt{smith02}) taken in twilight. The comparison star used (at 08:57:59.04 +03:41:54.5) has magnitudes of $u'=16.209$, $g'=14.997$, $r'=14.602$ and $i'=14.455$.  Using our extinction coefficients we extrapolated all fluxes to an airmass of $0$. The systematic error introduced by our flux calibration is $<0.1$ mag in all bands.

\subsection{VLT/X-shooter spectroscopy}

We obtained service mode observations of {\target} with X-shooter \citep{dodorico06} mounted at the VLT-UT2 telescope. The observations were designed to cover two entire orbits of the system. Details of these observations are listed in Table~\ref{obs}. X-shooter is a medium resolution spectrograph consisting of 3 independent arms that give simultaneous spectra longward of the atmospheric cutoff (0.3 microns) in the UV (the ``UVB'' arm), optical (the ``VIS'' arm) and up to 2.5 microns in the near-infrared (the ``NIR''arm). We used slit widths of 0.8'', 0.9'' and 0.9'' in X-shooter's three arms and $2\times2$ binning resulting in a resolution of $R\sim6,000$. After each exposure we nodded along the slit to help the sky subtraction in the NIR arm.

The reduction of the raw frames was conducted using the standard pipeline release of the X-shooter Common Pipeline Library (CPL) recipes (version 1.3.0) within ESORex, the ESO Recipe Execution Tool, version 3.9.0. The standard recipes were used to optimally extract and wavelength calibrate each spectrum. The instrumental response was removed by observing the spectrophotometric standard star EG 274 and dividing it by a flux table of the same star to produce the response function. We then heliocentrically corrected the wavelength scales of each of the spectra. For the NIR arm we combine frames taken at different nod positions to improve the sky subtraction. However, given the long exposure times in the NIR arm, this leads to a dramatic loss of orbital phase resolution. We achieve a signal-to-noise ($S/N$) of $\sim20$ in the UVB arm per exposure, $\sim10$ in the VIS arm per exposure and $\sim10$ in the NIR in two hours. The $K$ band data are completely dominated by sky noise. The S/N in the NIR arm is too low to detect any features from the secondary star.

The ULTRACAM and SOFI light curves were used to flux calibrate the X-shooter spectra. We fitted a model to each of the ULTRACAM light curves (see Section~\ref{lcurve}) and the SOFI $J$-band light curve in order to reproduce the light curve as closely as possible. The model was then used to predict the flux at the times of each of the X-shooter observations. We then derived synthetic fluxes from the spectra for the ULTRACAM $u'$, $g'$, $r'$, $i'$ and $z'$ filters as well as the SOFI $J$ and $H$ filters. We extrapolated the light curve models to those bands not covered by our photometry. We then calculated the difference between the model and synthetic fluxes and fitted a second-order polynomial to them. This correction was then applied to each spectrum. This corrects for variable extinction across the wavelength range, as well as variations in seeing.

\begin{figure*}
\begin{center}
 \includegraphics[width=0.8\textwidth]{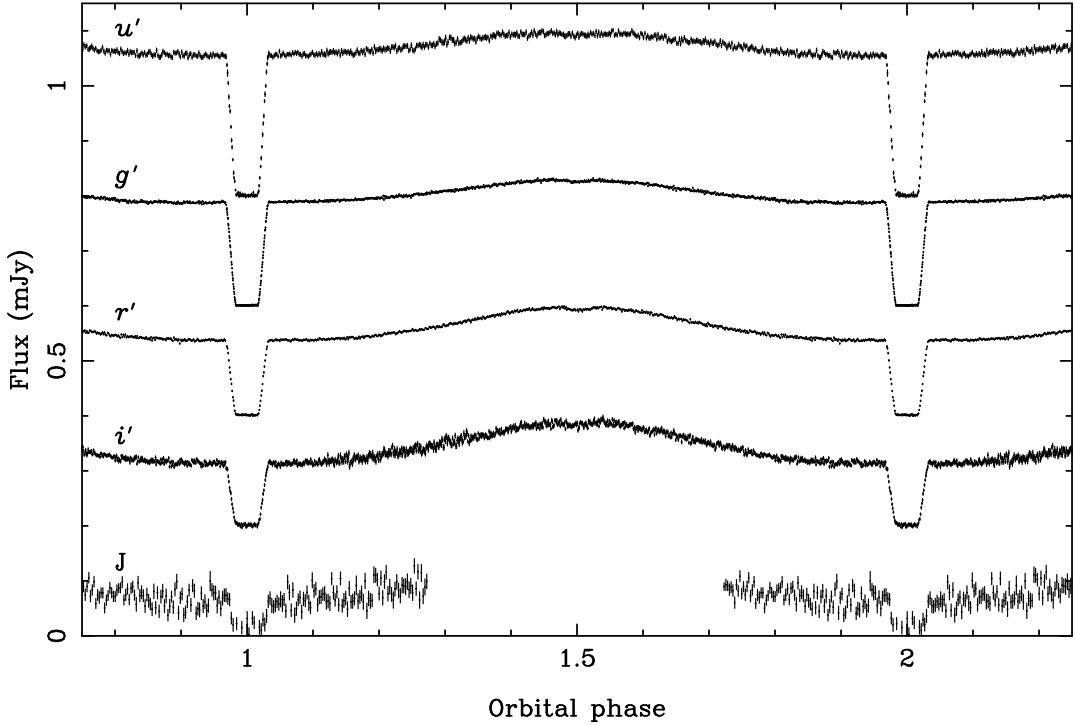}
 \caption{{\it Top to bottom:} phase-folded ULTRACAM $u'g'r'i'$ and SOFI $J$ band light curves of {\target}. The ULTRACAM light curves have been binned by a factor of three. Each light curve is offset by 0.2 mJy from the one below it. The $J$ band light curve has been scaled up by a factor of three for clarity.}
 \label{lcurves}
 \end{center}
\end{figure*}

\subsection{NTT/SOFI $J$ band photometry}

{\target} was observed with SOFI mounted at the NTT on the 6th April 2011. The observations were made in fast photometry mode equipped with a $J$ band filter. We windowed the detector to achieve a cycle time of 15 seconds and offset the telescope every 10 minutes in order to improve sky subtraction. 

Debiassing (which also removes the dark current) and flatfielding were performed in the standard way. Sky subtraction was achieved by using observations of the sky when the target had been offset. The average sky level was then added back so that we could determine the source flux and its uncertainty with standard aperture photometry, using a variable aperture, within the ULTRACAM pipeline. A comparison star (2MASS 08575095+0340301, $J=13.289$) was used to account for variations in observing conditions. Flux calibration was done using the comparison star $J$ band magnitude retrieved from the 2MASS catalogue \citep{skrutskie06}.

\section{Results}

\subsection{White dwarf temperature} \label{wdtemp}

\citet{eisenstein06} and \citet{drake10} both determined the temperature of the white dwarf in {\target} using its SDSS spectra. However, the reprocessed light from the heated face of the secondary star can lead to incorrect results from spectral fitting. Therefore, we combined the X-shooter spectra taken around the eclipse, where the reflection effect is minimal, and used this spectrum to determine the temperature and $\log{g}$ values using the method outlined in \citet{rebassa07}. Figure~\ref{specfit} shows the results of the spectral fit to the white dwarf features, we determined a temperature of $35,300\pm400$K and a $\log{g}$ of $7.3\pm0.1$, which is consistent with the light curve model fits and the fits to the SDSS spectra from \citet{eisenstein06} and \citet{drake10}. The fit also gave a distance to {\target} of $968\pm55$pc, consistent with previous measurements.

\subsection{Light curve model fitting} \label{lcurve}

\begin{figure*}
\begin{center}
 \includegraphics[width=0.7\textwidth]{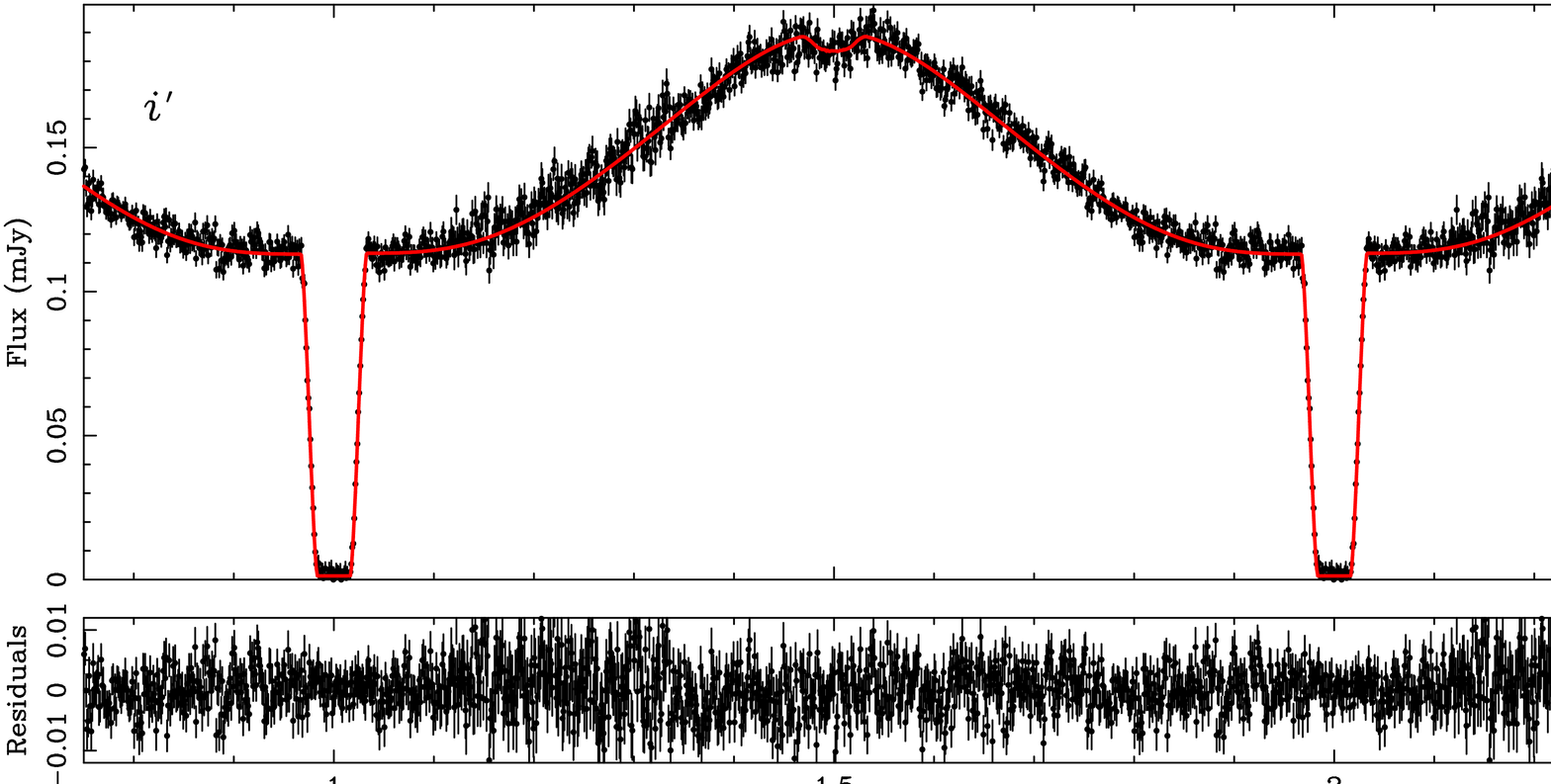}\\
 \vspace{4mm}
 \includegraphics[width=0.4\textwidth]{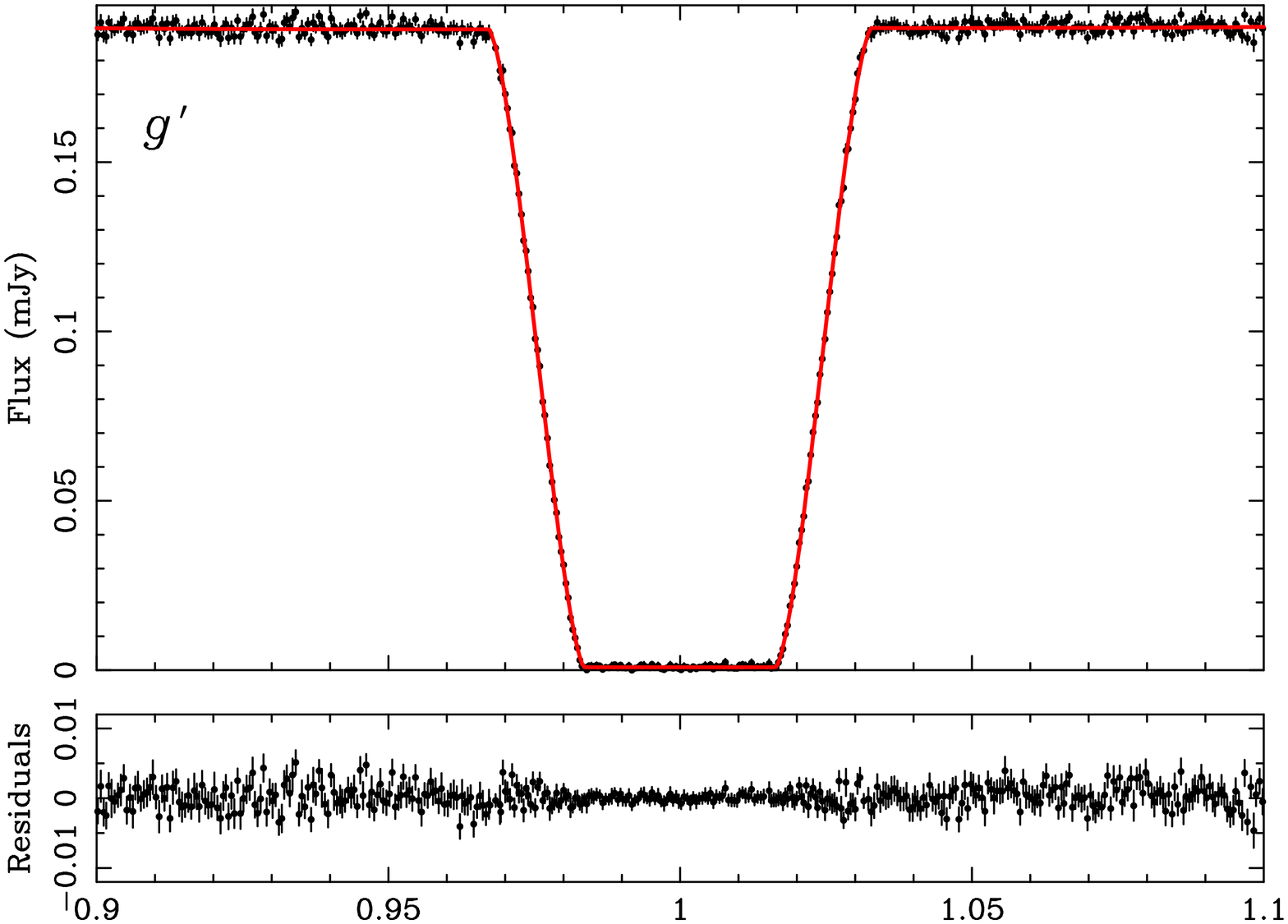}
 \hspace{1mm}
 \includegraphics[width=0.37\textwidth]{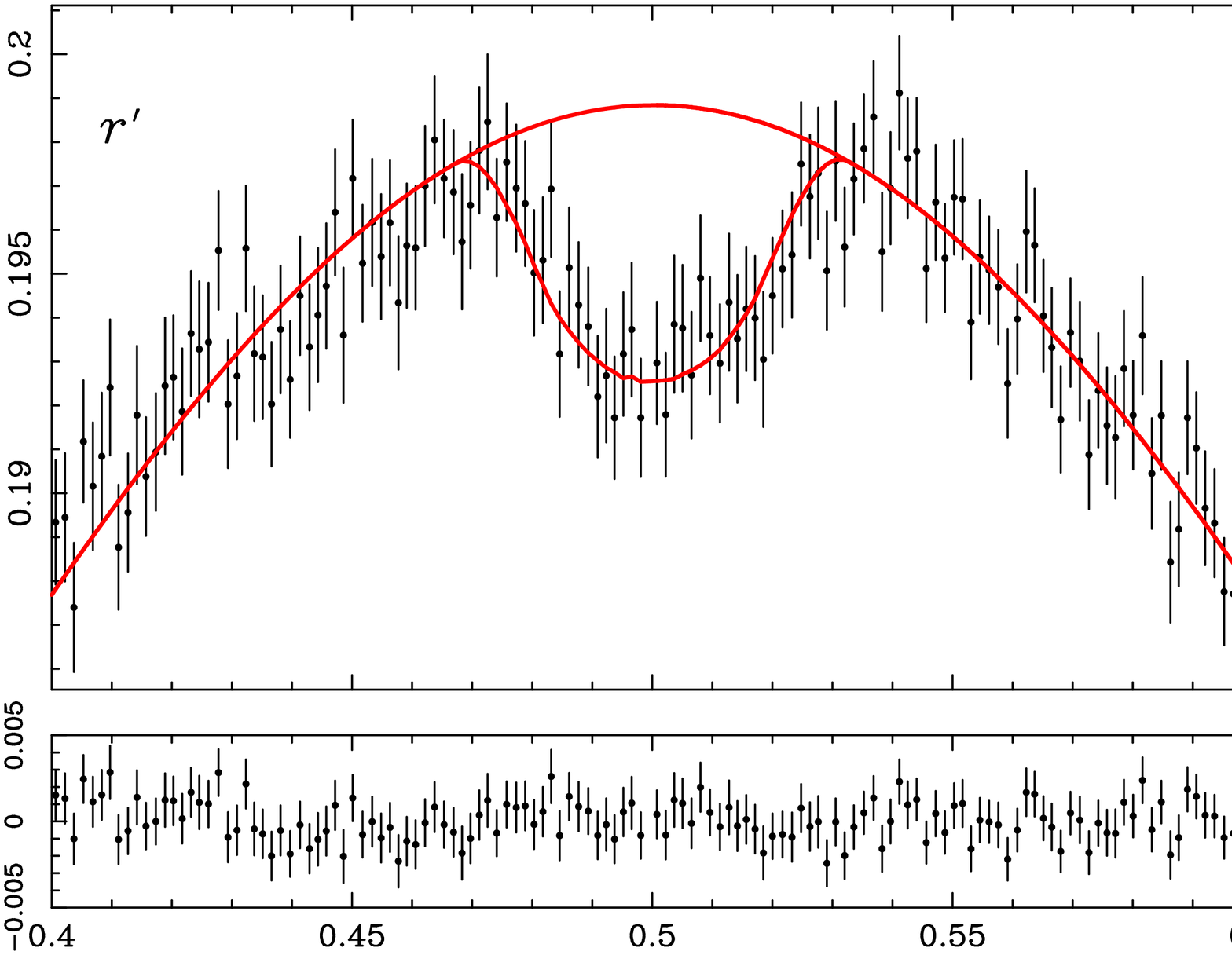}
 \vspace{2mm}
 \caption{Model fits to the ULTRACAM light curves with residuals shown below. \emph{Top}: Full orbit in the $i'$ band. \emph{Bottom left:} The $g'$ band primary eclipse. \emph{Bottom right:} The $r'$ band secondary eclipse, also shown is the same model but without the secondary eclipse.}
 \label{lcfitfig}
 \end{center}
\end{figure*}

Figure~\ref{lcurves} shows phase folded light curves for {\target} in the ULTRACAM $u'g'r'i'$ bands and the SOFI $J$ band. The deep primary eclipse is clearly visible in all bands; flux is detected in eclipse during the $J$ band eclipse only. A strong reflection effect is also evident in each of the bands caused by reprocessed light from the irradiated face of the secondary star. Also visible is the shallower secondary eclipse which becomes deeper in the redder bands (this phase was not covered by the SOFI observations). The detection of the secondary eclipse allows us to determine the inclination of the system and establish precise scaled radii, $R_\mathrm{sec}/a$ and $R_\mathrm{WD}/a$, where $a$ is the orbital separation.

\begin{table}
 \centering
  \caption{NTT/ULTRACAM mid-eclipse times for {\target}.}
  \label{ecl_times}
  \begin{tabular}{@{}lcc@{}}
  \hline
Cycle  & Eclipse time  & Uncert    \\ 
Number & MJD(BTDB)     & MJD(BTDB) \\
 \hline
-298   & 55533.3139971 & 0.0000024 \\
-237   & 55537.2848843 & 0.0000019 \\
-206   & 55539.3028807 & 0.0000024 \\
-191   & 55540.2793255 & 0.0000022 \\
-100   & 55546.2031193 & 0.0000064 \\
 -99   & 55546.2682076 & 0.0000020 \\
 240   & 55568.3359334 & 0.0000020 \\
 269   & 55570.2237344 & 0.0000024 \\
 299   & 55572.1766308 & 0.0000025 \\
\hline
\end{tabular}
\end{table}

To measure the system parameters we fitted the ULTRACAM light curves using a code written to produce models for the general case of binaries containing white dwarfs (see \citealt{copperwheat10} for details). It has been used in the study of other white dwarf-main sequence binaries (\citealt{pyrzas09}; \citealt{parsons10b}). The program subdivides each star into small elements with a geometry fixed by its radius as measured along the direction of centres towards the other star. Roche geometry distortion and irradiation of the secondary star are included, the irradiation is approximated by $\sigma T^{\prime}{}_\mathrm{sec}^{4}= \sigma T_\mathrm{sec}^{4}+ A F_\mathrm{irr}$ where $T^{\prime}{}_\mathrm{sec}$ is the modified temperature and $T_\mathrm{sec}$ is the temperature of the unirradiated main-sequence star, $\sigma$ is the Stefan-Boltzmann constant, $A$ is fraction of the irradiating flux from the white dwarf absorbed by the secondary star and $F_\mathrm{irr}$ is the irradiating flux, accounting for the angle of incidence and distance from the white dwarf. 

We used the Markov Chain Monte Carlo (MCMC) method to determine the distributions of our model parameters \citep{press07}. The MCMC method involves making random jumps in the model parameters, with new models being accepted or rejected according to their probability computed as a Bayesian posterior probability. In this instance this probability is driven by $\chi^2$ with all prior probabilities assumed to be uniform. A crucial practical consideration of MCMC is the number of steps required to fairly sample the parameter space, which is largely determined by how closely the distribution of parameter jumps matches the true distribution. We therefore built up an estimate of the correct distribution starting from uncorrelated jumps in the parameters, after which we computed the covariance matrix from the resultant chain of parameter values. The covariance matrix was then used to define a multivariate normal distribution that was used to make the jumps for the next chain. At each stage the actual size of the jumps was scaled by a single factor set to deliver a model acceptance rate of $\approx 25$ per cent. After 2 such cycles, the covariance matrix showed only small changes, and at this point we carried out the long ``production runs'' during which the covariance and scale factor which define the parameter jumps were held fixed. Note that the distribution used for jumping the model parameters does not affect the final parameter distributions, only the time taken to converge towards them.

The parameters needed to define the model were: the mass ratio, $q = M_\mathrm{sec}/M_\mathrm{WD}$, the inclination, $i$, the scaled radii $R_\mathrm{sec}/a$ and $R_\mathrm{WD}/a$, the unirradiated temperatures, $T_\mathrm{eff,WD}$ and $T_\mathrm{eff,sec}$, limb darkening coefficients for the both stars, the time of mid eclipse, $T_{0}$ and the period, $P$.

\begin{table*}
 \centering
  \caption{Parameters from Markov chain Monte Carlo minimisation, some fitted, some fixed a priori. $a_\mathrm{WD}$ and $b_\mathrm{WD}$ are the quadratic limb darkening coefficients of the white dwarf, $a_\mathrm{sec}$ is the linear limb darkening of the secondary star. $A$ is the fraction of the irradiating flux from the white dwarf absorbed by the secondary star.}
  \label{lcfit}
  \begin{tabular}{@{}lcccc@{}}
  \hline
 Parameter & $u'$ & $g'$ & $r'$ & $i'$ \\
 \hline 
$i$                 & $86.6\pm0.5$      & $85.4\pm0.2$      & $85.5\pm0.2$      & $84.8\pm0.5$       \\
$r_\mathrm{WD}/a$    & $0.0461\pm0.0012$ & $0.0438\pm0.0006$ & $0.0431\pm0.0005$ & $0.0423\pm0.0013$  \\
$r_\mathrm{sec}/a$   & $0.180\pm0.004$   & $0.192\pm0.003$   & $0.191\pm0.003$   & $0.203\pm0.008$    \\
$T_\mathrm{eff,sec}$ & $4922\pm97$       & $3575\pm67$       & $3298\pm64$       & $3016\pm104$       \\
$a_\mathrm{WD}$      & $0.089$           & $0.064$           & $0.057$           & $0.054$            \\
$b_\mathrm{WD}$      & $0.199$           & $0.178$           & $0.143$           & $0.119$            \\
$a_\mathrm{sec}$     & $-0.90\pm0.10$    & $-0.30\pm0.05$    & $-0.16\pm0.05$    & $-0.09\pm0.11$     \\
$A$                & $0.83\pm0.02$     & $0.64\pm0.02$     & $0.78\pm0.02$     & $0.69\pm0.06$      \\
 \hline
\end{tabular}
\end{table*} 

Each light curve was initially fitted in order to determine the mid-eclipse times and thus an accurate ephemeris. The mid-eclipse times are listed in Table~\ref{ecl_times}. We update the ephemeris to
\[\mathrm{MJD(BTDB)} = 55552.712\,765\,17(78) +\, 0.065\,096\,538(3) E,\]
which is consistent with the ephemeris of \citet{drake10}.

We then phase-folded the data and kept the period fixed but allowed the time of mid eclipse to vary. We kept the mass ratio fixed at 0.13 (the light curves are only weakly dependent on this parameter) and the temperature of the white dwarf fixed at $35,300$K (see section \ref{wdtemp}). We determined the quadratic limb darkening coefficients of the white dwarf based on a white dwarf atmosphere model with a temperature of $T_\mathrm{eff} = 35,300$K and $\log{g} = 7.3$ folded through the ULTRACAM $u'g'r'i'$ filter profiles \citep{gansicke95}. The coefficients ($a_\mathrm{WD}$ and $b_\mathrm{WD}$) for each filter are listed in Table~\ref{lcfit} for $I(\mu)/I(1) = 1-a(1-\mu)-b(1-\mu)^2$, where $\mu$ is the cosine of the angle between the line of sight and the surface normal; we kept these values fixed. All other parameters were allowed to vary and several chains were started with different initial parameters in order to check if the results were consistent and that we had found the global minimum.

The best fit parameters and their statistical errors are displayed in Table~\ref{lcfit}. Figure~\ref{lcfitfig} shows the fits to the light curves over the entire orbit as well as zoomed in around the two eclipses. The MCMC chains showed no variation beyond that expected from statistical variance. Since we didn't cover the secondary eclipse in the $J$ band and the $S/N$ of the SOFI data is quite low, we don't fit this data.

The four different bands give consistent results, although the $u'$ band favours a slightly higher inclination but is less constrained due to the shallowness of the secondary eclipse in this band. The faintness of the system in the $i'$ band limits the constraints in this band. We found an inclination of $85.5^{\circ}\pm0.2^{\circ}$ and scaled radii of $R_\mathrm{WD}/a = 0.0436\pm0.0004$ and $R_\mathrm{sec}/a = 0.190 \pm 0.002$. Given our black body assumption, $T_\mathrm{sec}$ only approximately represents the true temperature of the secondary star, its more important role is as a flux scaling factor. An interesting trend is seen in the limb darkening coefficients for the secondary star, which are all negative (limb brightening) and the amount of limb brightening decreases with increasing wavelength. This is presumably the result of seeing to different depths at different wavelengths; a similar effect was seen with NN Ser \citep{parsons10}.

\subsection{Radial Velocities}

\begin{figure*}
\begin{center}
 \includegraphics[width=0.8\textwidth]{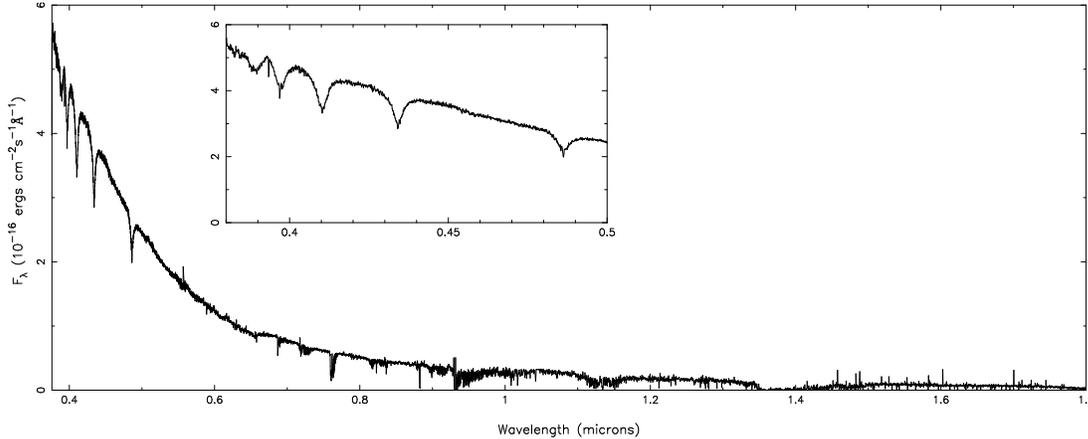}
 \vspace{2mm}
 \caption{Averaged X-shooter spectrum of {\target}. No features from the secondary star are seen in the NIR, the only features visible are telluric or sky lines that have not been completely removed. Those spectra taken during the eclipse were not included in the averaging. Inset is a zoom in on the upper Balmer series, interstellar calcium absorption is also visible.}
 \label{av_spec}
 \end{center}
\end{figure*}

\begin{figure*}
\begin{center}
 \includegraphics[width=0.8\textwidth]{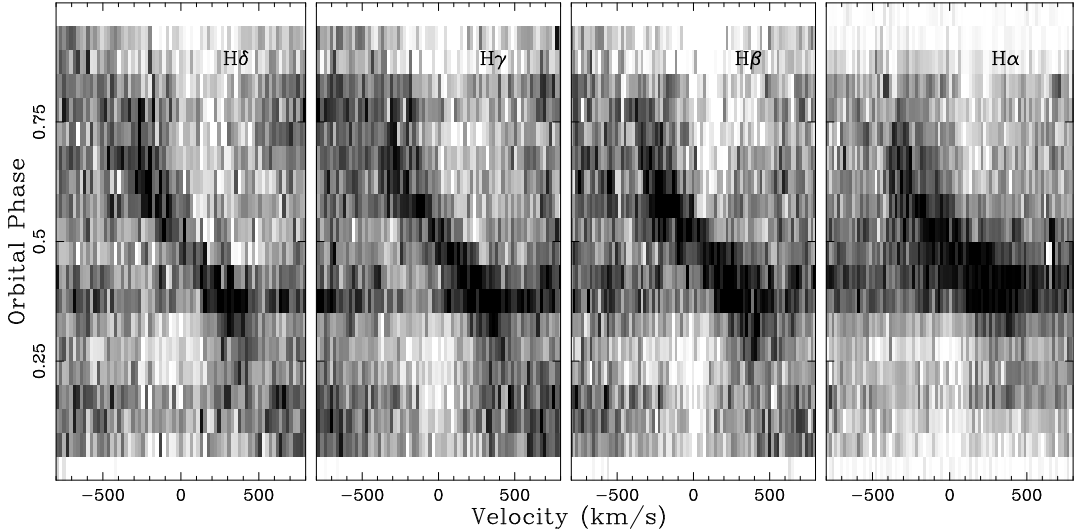}
 \vspace{-2mm}
 \caption{Trailed spectrum of the four longest wavelength hydrogen Balmer lines, the scale runs from white (75 per cent of the continuum level) to black (120 per cent of the continuum level). Both white dwarf absorption and secondary star emission components are visible in each line.}
 \label{trails}
 \end{center}
\end{figure*}

Figure~\ref{av_spec} shows an average spectrum of {\target}. The spectral features seen are similar to other PCEBs, namely hydrogen Balmer absorption originating from the white dwarf and emission from the heated face of the secondary star, which disappears around the primary eclipse. Calcium emission is also seen as well as deep interstellar calcium H and K absorption which show no radial velocity variations; no other features from either star are visible. Figure~\ref{trails} shows trailed spectra of the four longest wavelength hydrogen Balmer lines. The emission from the secondary star can be seen as well as the non-LTE absorption core from the white dwarf moving in anti-phase.

\begin{table}
 \centering
  \caption{Hydrogen Balmer line offsets and velocities.}
  \label{lines}
  \begin{tabular}{@{}lcccc@{}}
  \hline
Line  & $\gamma_\mathrm{WD}$ & $K_\mathrm{WD}$  & $\gamma_\mathrm{emis}$ & $K_\mathrm{emis}$ \\
      & (km$\,$s$^{-1}$)    & (km$\,$s$^{-1}$) &(km$\,$s$^{-1}$)       &(km$\,$s$^{-1}$)  \\
 \hline
H $\delta$ & $-$          & $-$           & $28.9\pm4.9$ & $319.4\pm8.7$ \\
H $\gamma$ & $48.4\pm8.6$ & $65.9\pm11.8$ & $25.5\pm5.3$ & $324.7\pm8.9$ \\
H $\beta$  & $50.6\pm5.8$ & $61.6\pm8.8$  & $28.5\pm5.3$ & $343.6\pm9.8$ \\
H $\alpha$ & $42.1\pm6.3$ & $66.7\pm10.0$ & $31.8\pm5.0$ & $336.2\pm9.9$ \\
\hline
\end{tabular}
\end{table}

For each of the Balmer lines we fit both the absorption and emission components together by simultaneously fitting all of the spectra. We used this approach because the $S/N$ on an individual spectrum was quite low and the absorption from the white dwarf is filled in by the emission component from the secondary star around phase 0.5. Therefore, by fitting all of the spectra simultaneously we obtain a more robust estimate of the radial velocities. We used a combination of a straight line and Gaussians for each spectrum (including a broad Gaussian component to account for the wings of the primary star's absorption) and allowed the position of the Gaussians to change velocity according to
\begin{eqnarray}
V = \gamma + K\sin(2 \pi \phi), \nonumber
\end{eqnarray}
for each star, where $\gamma$ is the velocity offset of the line from its rest wavelength and $\phi$ is the orbital phase of the spectrum. We allow the height of the emission component to vary with orbital phase according to
\begin{eqnarray}
H = H_0 - H_1\cos(2 \pi \phi), \nonumber
\end{eqnarray}
which allows the height to peak at phase 0.5, where the irradiation effect is largest. This approach gave better fits than keeping the height at a fixed value.

Table~\ref{lines} lists the offsets and radial velocities for the four longest wavelength hydrogen Balmer lines, the shorter wavelength lines were unsuitable for fitting since the emission component is much weaker and the absorption components lack the non-LTE core, this is also the case for the H $\delta$ absorption. The results for each line are consistent and we adopt values of $K_\mathrm{WD} = 64\pm6$ km$\,$s$^{-1}$ and $K_\mathrm{emis}=330\pm5$ km$\,$s$^{-1}$. We also measure a gravitational redshift for the white dwarf of $z_\mathrm{WD} = \gamma_\mathrm{WD} - \gamma_\mathrm{emis} = 18\pm5$ km$\,$s$^{-1}$.

\subsection{$K_\mathrm{sec}$ correction}

The emission lines seen in the X-shooter spectra are the result of reprocessed light from the surface of the secondary star facing the white dwarf, hence their radial velocity amplitude represents a lower limit to the true centre of mass radial velocity amplitude. For accurate mass determinations the centre of mass radial velocity amplitude is required, thus we need to determine the deviation between the reprocessed light centre and the centre of mass for the secondary star. The radial velocity of the centre of mass ($K_\mathrm{sec}$) is related to that of the emission lines ($K_\mathrm{emis}$) by
\begin{eqnarray}
K_\mathrm{sec} = \frac{K_\mathrm{emis}}{1 - f(1+q)\frac{R_\mathrm{sec}}{a}}
\end{eqnarray}
where $f$ is a constant between 0 and 1 which depends upon the location of the centre of light. For $f=0$ the emission is spread uniformly across the entire surface of the secondary star and therefore the centre of light is the same as the centre of mass. For $f=1$ all of the flux is assumed to come from the point on the secondary star's surface closest to the white dwarf (the substellar point). The maximum possible value for $K_\mathrm{sec}$ is therefore 423 km$\,$s$^{-1}$ (using $K_\mathrm{emis}=330$ km$\,$s$^{-1}$, $R_\mathrm{sec}/a=0.190$ and $f=1$). 

The centre of light for an emission line is related to the optical depth of the emission \citep{parsons10}. Optically thick emission tends to be preferentially radiated perpendicular to the stellar surface, therefore at the quadrature phases, we will see the limb of the irradiated region more prominently (compared to the region of maximum irradiation) than we would otherwise. This will lead to a higher observed semi-amplitude and hence a smaller correction factor is needed. The reverse is true for optically thin lines where the emission is radiated equally in all directions, hence emission from the substellar point becomes enhanced at quadrature, leading to a low semi-amplitude and a larger correction factor. The correction factor for an optically thin line (assuming emissivity proportional to the incident flux) is $f=0.77$ \citep{parsons10}. Therefore, a more stringent upper limit on the correction factor can be set by using an optically thin correction since this represents a more physical upper limit. This gives a value of $K_\mathrm{sec,max}=398$ km$\,$s$^{-1}$ (using $f=0.77$). The hydrogen Balmer emission is likely to be optically thick, as found for the similarly irradiated secondary star in the PCEB NN Ser \citet{parsons10}. We will show in Section~\ref{sys_paras} it is the lower limit that is the more important. The $S/N$ of the X-shooter data is too low to carry out a similar analysis to that presented in \citet{parsons10} for NN Ser. A lower limit can be obtained by assuming that the emission is uniformly spread over the heated face (in reality it is far more likely to be brighter towards the substellar point where the heating effect is larger). This corresponds to a correction factor of $f=0.42$ which gives $K_\mathrm{sec,min}=364$ km$\,$s$^{-1}$. 

Therefore, with the current data we are only able to say that the radial velocity semi-amplitude of the centre of mass of the secondary star ($K_\mathrm{sec}$) lies somewhere between 364 km$\,$s$^{-1}$ and 398 km$\,$s$^{-1}$.

\section{Discussion}

\begin{table}
 \centering
  \caption{Stellar and binary parameters for {\target}. The white dwarf temperature and $\log{g}$ are from the spectroscopic fit. $z_\mathrm{WD}$ is the measured gravitational redshift of the white dwarf.}
  \label{params}
  \begin{tabular}{@{}lccc@{}}
  \hline
  Parameter & Value         & Parameter & Value \\
  \hline
  RA     & 08:57:46.18      & P$_\mathrm{orb}$ (days)        & 0.065\,096\,538(3) \\
  Dec    & +03:42:55.3      & T$_\mathrm{WD}$ (K)            & $35,300\pm400$ \\
  $u'$   & $17.731\pm0.011$ & WD $\log{g}$                  & $7.3\pm0.1$ \\
  $g'$   & $17.954\pm0.006$ & M$_\mathrm{WD}$ (M$_{\sun}$)    & $0.514\pm0.049$ \\
  $r'$   & $18.256\pm0.007$ & M$_\mathrm{sec}$ (M$_{\sun}$)   & $0.087\pm0.012$ \\
  $i'$   & $18.393\pm0.011$ & R$_\mathrm{WD}$ (R$_{\sun}$)    & $0.0247\pm0.0008$ \\
  $z'$   & $18.536\pm0.032$ & R$_\mathrm{sec}$ (R$_{\sun}$)   & $0.1096\pm0.0038$ \\
  $J$    & $18.184\pm0.059$ & a (R$_{\sun}$)                 & $0.574\pm0.018$ \\
  $H$    & $18.618\pm0.160$ & $i$ (deg)                      & $85.5\pm0.2$ \\
  $K$    & $18.141\pm0.174$ & $K_\mathrm{WD}$ (km$\,$s$^{-1}$) & $64\pm6$ \\
  d (pc) & $968\pm55$       & $K_\mathrm{sec}$ (km$\,$s$^{-1}$) & $364<K<398$ \\
  Sp2    & M$8\pm1$         & $z_\mathrm{WD}$ (km$\,$s$^{-1}$) & $18\pm5$ \\
\hline
\end{tabular}
\end{table}

\subsection{System parameters} \label{sys_paras}

We can combine our light curve model fits, our radial velocity measurements and Kepler's third law to determine the masses, radii and orbital separation for {\target}. These are listed in Table~\ref{params} along with the other system parameters that we have measured, following the procedure outlined in \citet{parsons10}.

\begin{figure}
\begin{center}
 \includegraphics[width=\columnwidth]{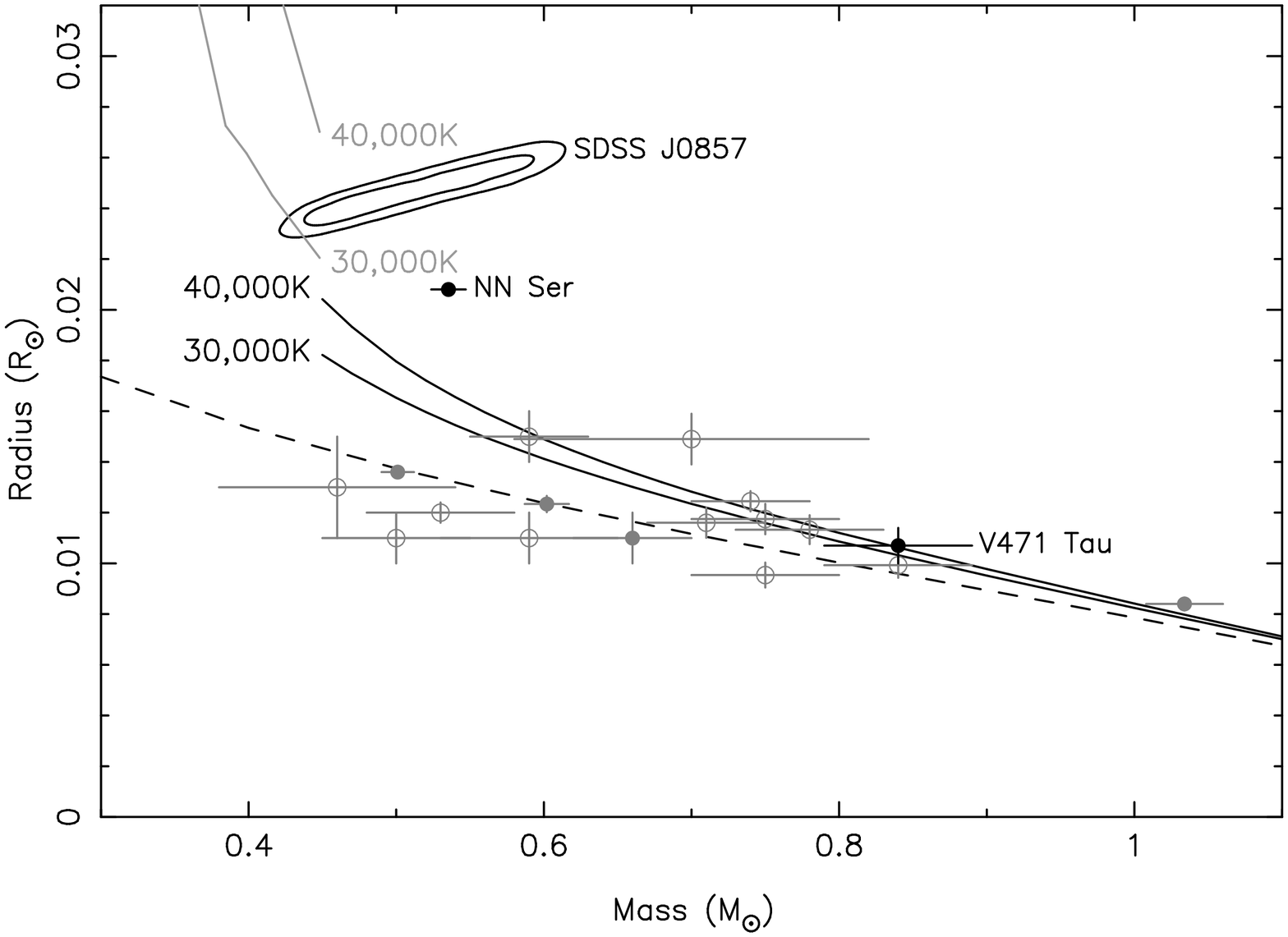}
 \caption{Mass-radius plot for white dwarfs. Data from \citet{provencal98}, \citet{provencal02} and \citet{casewell09} are plotted. The filled circles are visual binaries and the open circles are common proper-motion systems. The solid dots are from the eclipsing PCEBs NN Ser \citep{parsons10} and V471 Tau \citep{obrien01}. The solid black lines correspond to carbon-oxygen core models with temperatures of 30,000K and 40,000K with a hydrogen layer thickness of M$_\mathrm{H}/\mathrm{M}_\mathrm{WD} = 10^{-4}$ from \citet{benvenuto99}. The solid grey lines correspond to helium core models with temperatures of 30,000K and 40,000K from \citet{panei07}. The dashed line is the zero-temperature mass-radius relation of Eggleton from \citet{verbunt88}. The contours show the 68 and 95 percentile regions of the masses and radii for the white dwarf in {\target}. For the measured temperature of the white dwarf (35,300K) the measured radius is much larger than predicted from the carbon-oxygen core models but is consistent with the helium core models.}
 \label{wd_MR}
 \end{center}
\end{figure}

The largest source of uncertainty is $K_\mathrm{sec}$ since we were unable to measure it directly. This translates into a large uncertainty in the masses (particularly the white dwarf). The contours in Figure~\ref{wd_MR} show the region of allowed mass and radius for the white dwarf in {\target} compared to other white dwarfs with measured masses and radii. Also plotted are mass-radius relations for carbon-oxygen core white dwarfs from \citet{benvenuto99} and helium core white dwarfs from \citet{panei07} in the same temperature range as the white dwarf in {\target}. The mass range permitted makes it possible for the white dwarf to have either type of core but the measured radius of the white dwarf places it far above the carbon-oxygen core models, even for very thick hydrogen envelopes. However, it is consistent with a helium core provided that the mass is at the lower end of the permitted range. Therefore, it is most likely that the white dwarf in {\target} is a helium core white dwarf close to the upper mass limit expected from evolution ($<0.5$M$_{\sun}$).

\begin{figure}
\begin{center}
 \includegraphics[width=\columnwidth]{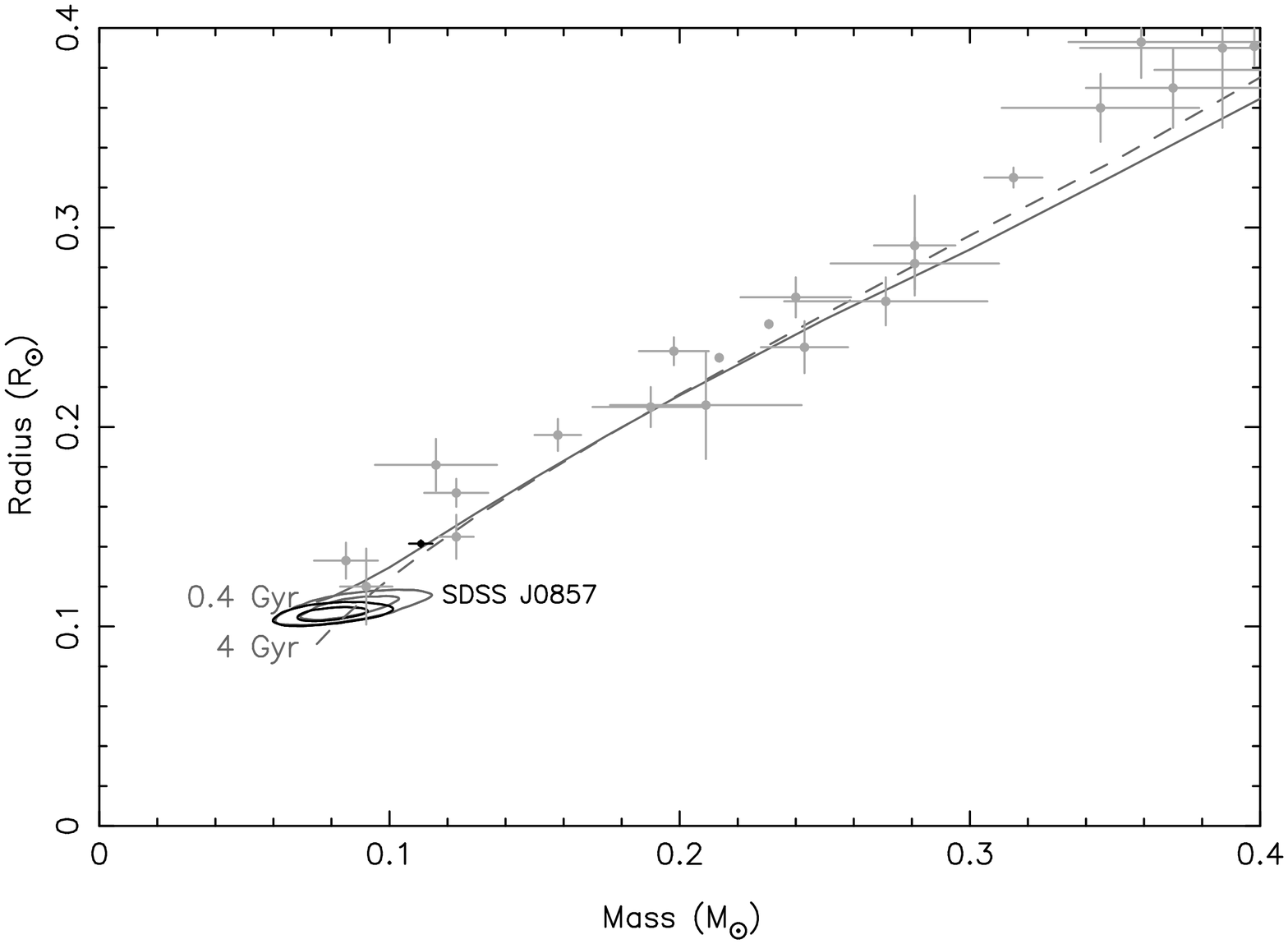}
 \caption{Mass-radius plot for low mass stars. Data from \citet{lopez07} and \citet{beatty07}, also plotted is the mass-radius measurement for the secondary star in the eclipsing PCEB NN Ser (black point; \citealt{parsons10}). The solid line represents the theoretical isochrone model from \citet{baraffe98}, for an age of 0.4 Gyr, solar metalicity, and mixing length $\alpha = 1.0$, the dashed line is the same but for an age of 4 Gyr. The grey contours show the 68 and 95 percentile regions of the masses and radii for the secondary star in {\target}. The black contours show the same regions if the white dwarf mass is forced to be $<0.5$M$_{\sun}$.}
 \label{sec_MR}
 \end{center}
\end{figure}

The contours in Figure~\ref{sec_MR} show the region of allowed mass and radius for the secondary star in {\target}. The mass range places it close to the hydrogen burning limit and the radius is consistent with the evolutionary models of \citet{baraffe98} despite being both heavily irradiated by the nearby white dwarf and rapidly rotating. This result is consistent with that found by \citet{parsons10} for the eclipsing PCEB NN Ser, implying that very low-mass stars do not appear to be overinflated compared to evolutionary models, in contrast to slightly more massive stars ($\ga$0.3M$_{\sun}$) \citep{kraus11}. The black contours in Figure~\ref{sec_MR} show the region of allowed mass and radius if the white dwarf mass is $<0.5$M$_{\sun}$ (a helium core), in this case the mass and radius are slightly smaller ($0.08\pm0.01$M$_{\sun}$ and $0.106\pm0.002$R$_{\sun}$).

The secondary star was detected during the SOFI $J$ band eclipse and we measure its magnitude as $J_\mathrm{sec} = 20.9\pm0.2$ which, at the distance determined for the white dwarf, gives an absolute magnitude of $M_{J,\mathrm{sec}} = 11.0\pm0.2$. This corresponds to a spectral type of M$8\pm1$ \citep{hawley02} and is consistent with the measured mass \citep{baraffe96}.

Using the X-shooter spectra, the gravitational redshift of the white dwarf was found to be $18\pm5$ km$\,$s$^{-1}$. Using the measured mass and radius from Table~\ref{params} gives a redshift of $13.2$ km$\,$s$^{-1}$. Correcting the X-shooter measurement for the redshift of the secondary star, the difference in transverse Doppler shifts and the potential at the secondary star owing to the white dwarf gives a value of $12\pm1$ km$\,$s$^{-1}$ which is consistent with the measured value.

\subsection{Evolutionary state}

Since {\target} has emerged from the common envelope stage its evolution has been driven by angular momentum loss, which will eventually bring the two stars close enough to initiate mass transfer from the secondary star to the white dwarf. For short period binaries angular momentum can be lost via gravitational radiation (\citealt{kraft62}; \citealt{faulkner71}) and a magnetised stellar wind, known as magnetic braking (\citealt{verbunt81}; \citealt{rappaport83}). Magnetic braking becomes much weaker in binaries with very low mass stars ($M\la0.3 M_{\sun}$) since they become fully convective and the magnetic field is no longer rooted to the stellar core. However, residual magnetic braking in CVs with fully convective donor stars has been suggested both based on the location of the observed orbital period minimum (\citealt{patterson98}; \citealt{kolb99}) and on the differences in the temperatures of the accretion-heated white dwarfs in strongly magnetic and non-magnetic CVs \citep{townsley09}. Independent evidence for magnetic braking comes from studies of single low mass stars \citep{reiners08}.

To determine the past and future evolution of {\target} we interpolate the cooling tracks of \citet{panei07} for a helium core white dwarf with a mass of 0.45 M$_{\sun}$ and estimate its cooling age to be $\sim19.5\times10^6$ yrs, implying that the system has only recently emerged from the common envelope phase at essentially the current orbital period. An upper limit on the time taken for the system to reach a semi-detached configuration can be made by assuming that the only mechanism of angular momentum loss is via gravitational radiation. If this is the case then mass transfer will begin in $\sim3.7\times10^8$ years \citep{schreiber03} once the orbital period has been reduced to 66 minutes \citep{ritter86}. Since there is likely to be some additional angular momentum loss besides gravitational radiation, the true time taken to reach a semi-detached configuration is likely to be lower. {\target} will start mass transfer below the observed orbital period minimum for cataclysmic variables (CVs), $\sim80$ minutes, at which point the white dwarf will have cooled to around 13,500K. 

Once the binary becomes mass-transferring the secondary star will no longer be in thermal equilibrium and the binary will evolve to longer periods and pass into the ``main'' CV population. Therefore, {\target} is a progenitor for a class of CVs yet to be unambiguously observed: those that began mass transfer from a very low-mass star / brown dwarf donor. The few (non-AM CVn) CVs observed with periods below the minimum were not formed directly from white dwarf / brown dwarf binaries; V485 Cen ($P_\mathrm{orb}=59$ minutes) and EI Psc ($P_\mathrm{orb}=64$ minutes) are both thought to have evolved donor stars in which a large fraction of the hydrogen in the core was processed prior to mass transfer (\citealt{thorstensen02}; \citealt{gansicke03}). SDSS\,J1507 ($P_\mathrm{orb}=66$ minutes) was thought to have started mass transfer with a brown dwarf donor \citep{littlefair07} but recent studies have shown that it is more likely to be a halo CV with a low-metalicity donor (\citealt{patterson08}; \citealt{uthas11}). Once it has reached a semi-detached state, {\target} is likely to take about a Gyr to evolve back towards a period where the shortest period ``main'' CVs are found \citep{kolb99}.

Our constraints on the system parameters of {\target} make it likely that the progenitor for this system lied within, or close to, the brown dwarf desert; a paucity of close ($\la$5AU) brown dwarf companions to main-sequence stars \citep{grether06}.

\section{Conclusions}

We have used a combination of ULTRACAM and SOFI photometry and X-shooter spectroscopy to constrain the system parameters of the 94 minute orbital period eclipsing PCEB {\ltarget}. We measure a temperature for the white dwarf of $35,300\pm400$K and a distance of $968\pm55$pc. By detecting the secondary eclipse in our ULTRACAM photometry we were able to determine an inclination of $i=85.5^{\circ}\pm0.2^{\circ}$. We were also able to measure the radial velocity amplitude of the white dwarf as $64\pm6$ km$\,$s$^{-1}$. No absorption features from the secondary star were visible in the X-shooter spectra, however we were able to determine the range of possible $K_\mathrm{sec}$ values by applying a correction factor to the measured radial velocity amplitude of several emission lines originating from its heated face. 

By combining the radial velocity measurements with the light curve fit we measured the mass and radius of the white dwarf to be $0.51\pm0.05$M$_{\sun}$ and $0.0247\pm0.0008$R$_{\sun}$ respectively. The measured temperature, mass and radius are inconsistent with a carbon-oxygen core white dwarf but are consistent with a helium core white dwarf at the upper mass range for helium core white dwarfs. The secondary star has a mass of $0.09\pm0.01$M$_{\sun}$ and a radius of $0.110\pm0.004$ consistent with evolutionary models of low-mass stars. Its mass places it at the hydrogen burning limit.

We also find that {\target} has only recently emerged from the common envelope phase and will reach a semi-detached configuration in $\sim4\times10^8$ years when it will become a cataclysmic variable with a 66 minute orbital period, at which point its orbital period will increase.

Due to the detection of the secondary eclipse, {\target} is ideally set up to measure precise masses and radii for both of the stars. Currently the limiting factor is the poorly constrained $K_\mathrm{sec}$. With higher $S/N$ spectra a more reliable $K$-correction can be made, or potentially a direct measurement via absorption lines in the infrared, which will vastly improve the precision of the masses and allow us to test the mass radius relation for massive helium core white dwarfs and very low-mass stars.

\section*{Acknowledgements}
ULTRACAM, TRM, BTG, CMC, VSD and SPL are supported by the Science and Technology Facilities Council (STFC). SPL also acknowledges the support of an RCUK Fellowship. ARM acknowledges financial support from FONDECYT in the form of grant number 3110049. MRS thanks for support from FONDECYT (1100782). The results presented in this paper are based on observations collected at the European Southern Observatory under programme IDs 086.D-0265, 286.D-5030 and 087.D-0046. 

\bibliographystyle{mn2e}
\bibliography{eclipsers}

\label{lastpage}

\end{document}